\documentclass[preprint,review,12pt,sort&compress]{elsarticle} 

\journal{Physica A}

\usepackage{bm}
\usepackage{amsmath}
\usepackage{bm}
\usepackage{xspace}

\begin{document}

\begin{frontmatter}



\title{Equilibration in finite Bose systems}


\author{Georg~Wolschin\corref{cor}}
\ead{g.wolschin@thphys.uni-heidelberg.de}

\address{Institut f{\"ur} Theoretische Physik der Universit{\"a}t Heidelberg, Philosophenweg 16, D-69120 Heidelberg, Germany, EU}

\cortext[cor]{Corresponding author}

\begin{abstract}
The equilibration of a finite Bose system is modeled using a gradient expansion of the collision integral
that leads to a nonlinear transport equation. For constant transport coefficients, it is solved in closed form through a nonlinear transformation.
Using schematic initial conditions, the exact solution and the equilibration time are derived and 
compared to the corresponding case for fermions.
Applications to the fast equilibration of the gluon system created initially in relativistic heavy-ion collisions, 
and to cold quantum gases are envisaged.
\end{abstract}


\end{frontmatter}
\newpage

\section{Introduction}

The analytical modeling of statistical equilibration processes in finite Bose systems is a challenging problem.
Here, condensation can accompany boson thermalization
such that the final state may deviate from the purely thermal distribution \cite{setk95,sto91,jpb12}.
In Fermi systems, the physics of thermalization is very different since Pauli's exclusion
principle determines the structure of
the final state \cite{gw82}. 

The Boltzmann equation with the corresponding statistical factors in the collision term
provides a starting point to model the equilibration in a system of fermions or bosons. 
It has been particularly useful to describe relativistic heavy-ion collisions, following early 
works to treat the collision term in the relaxation-time approximation \cite{bay84}
or with gluon--gluon scattering \cite{mue00}, and later numerical simulations.
Analytical solutions of the relativistic Boltzmann equation can lead to valuable
insights into nonlinear phenomena \cite{baz16}.

In this work the aim is to analytically solve a nonlinear kinetic equation
for the time evolution of the occupation-number distribution in a finite system of bosons 
that preserves the essential features of Bose--Einstein
statistics which are contained in the Boltzmann equation.

\section{Boson equilibration model}

Assuming spatial homogeneity for the boson distribution function $f(\bm{x},\bm{p},{t})$\, with momentum $\bm{p}$\,
or energy $\epsilon\,(\bm{p})$\, and a spherically symmetric momentum dependence, one can reduce
the kinetic equations to one dimension by carrying out the angular integration \cite{fuku17}. The equation for the
single-particle occupation numbers $n_j\equiv\,n_\text{th}(\epsilon_j,t)$ in a Bose system becomes
\begin{eqnarray}
\frac{\partial n_1}{\partial t}=\sum_{\epsilon_2,\epsilon_3,\epsilon_4}^\infty  \langle V_{1234}^2\rangle\,G\,(\epsilon_1+\epsilon_2, \epsilon_3+\epsilon_4)\times\qquad\qquad\\ \nonumber
\bigl[(1+n_1)(1+n_2)\,n_3\,n_4-
(1+n_3)(1+n_4)\,n_1\,n_2\bigr]
 \label{boltzmann}
\end{eqnarray}
with the second moment of the interaction $\langle V^2\rangle$ and the energy-conserving function $G$. 
It is a $\delta$-function for an infinite system as in the usual Boltzmann collision term
where the single-particle energies are time independent,\begin{equation}
G\,(\epsilon_1+\epsilon_2, \epsilon_3+\epsilon_4)\rightarrow \pi\, \delta(\epsilon_1+\epsilon_2- \epsilon_3-\epsilon_4)\,.
 \label{delta}
\end{equation}
With an energy-conserving $\delta$-function the system shows detailed balance and particle number is conserved when
summing over all energies $\epsilon \ge 0$. 

In a finite system the  energy-conserving function may, however, acquire a width
such that off-shell scatterings between single-particle states which lie apart in energy space become possible \cite{gww81}.

In both cases the collision term can also be written
 in the form of a master
equation with gain and loss terms, respectively
\begin{equation}
\frac{\partial n_1}{\partial t}=(1+n_1)\sum_{\epsilon_4} W_{4\rightarrow1}\,n_4-n_1\sum_{\epsilon_4}W_{1\rightarrow4}(1+n_4) 
 \label{boltz}
\end{equation}
with the transition probability
\begin{equation}
W_{4\rightarrow1}=\sum_{\epsilon_2,\epsilon_3}\, \langle V_{1234}^2\rangle\,G\,(\epsilon_1+\epsilon_2, \epsilon_3+\epsilon_4)\,(1+n_2)\,n_3
 \label{trans}
\end{equation}
and $W_{1\rightarrow 4}$ accordingly. The summations are then replaced by integrations, introducing the densities of states $g_j\equiv g(\epsilon_j)$
and $W_{4\rightarrow 1}=W_{41}g_1, W_{1\rightarrow 4}=W_{14}g_4$.
Because bosons are interchangeable, we have $W_{41}=W_{14}$.

If the energy-conserving function acquires a width in a finite system as discussed above, $W_{14}=W_{41}=W[\frac{1}{2}(\epsilon_4+\epsilon_1),|\epsilon_4-\epsilon_1|]$
depends on the absolute value of $x = \epsilon_4 - \epsilon_1$ and is peaked at $x=0$.
It is nonlinear because of the dependence on the statistical factors in Eq.\,({\ref{trans}), 
but here I shall explicitly treat only the nonlinearity in Eq.\,({\ref{boltz}) 
with the option to improve this by an iteration scheme. 

An approximation to Eq.\,({\ref{boltz}) can then be obtained through a gradient expansion in energy space of $n_4$ and $g_4\,n_4$
around $\epsilon_4=\epsilon_1$ to second order. By introducing transport coefficients via moments of the transition probability
\begin{eqnarray}
D=\frac{1}{2}\,g_1\int_0^\infty W(\epsilon_1,x)\,x^2dx,~~
 v=g_1^{-1}\frac{d}{d\epsilon_1}(g_1D)
 \label{moments}
\end{eqnarray}
one arrives at a nonlinear partial differential equation for $n\equiv n(\epsilon_1,t)=n\,(\epsilon,t)$
 \begin{equation}
\frac{\partial n}{\partial t}=-\frac{\partial}{\partial\epsilon}\bigl[v\,n\,(1+n)-n^2\frac{\partial D}{\partial \epsilon}\bigr]+\frac{\partial^2}{\partial\epsilon^2}\bigl[D\,n\bigr]\,.
 \label{boseq}
\end{equation}\\
Dissipative effects are expressed through the drift term $v(\epsilon,t)$, diffusive effects through the diffusion term $D(\epsilon,t)$.
In the limit of constant transport coefficients, the nonlinear boson diffusion equation  
for the occupation-number distribution $n\,(\epsilon,t)$
becomes
\begin{equation}
\frac{\partial n}{\partial t}=-v\,\frac{\partial}{\partial\epsilon}\bigl[n\,(1+n)\bigr]+D\,\frac{\partial^2n}{\partial\epsilon^2}\,,
 \label{bose}
\end{equation}
and the aim is to solve this kinetic equation exactly.
The thermal distribution is a stationary solution 
\begin{equation}
n_\text{eq}(\epsilon)=\frac{1}{e^{(\epsilon-\mu)/T}-1}
 \label{Bose--Einstein}
\end{equation}
with the chemical potential $\mu<0$ in a finite boson system.
The equilibrium temperature is given in terms of the transport coefficients, $T=-D/v$ with $v<0$ since the drift is towards the infrared region. 
An equilibrium distribution with $\mu=0$, and a constant distribution also solve Eq.\,(\ref{bose}).  The particle content in the condensate \cite{sto91,setk95,jpb12} is $(2\pi)^3\,n_\text{c}(t)\times\delta(\bm{p})$, 
with $n_\text{c}(t)$ the number density of bosons in the condensed state. 

The present analytical model does not treat the second-order phase transition to the condensate
below a critical temperature $T_c$ explicitly since a Boltzmann-type approach cannot account for the buildup of coherence which is required for the phase transition to occur \cite{sto91}. The model does treat the kinetics of Bose condensation before and after the phase transition as in related numerical approaches \cite{setk95,setk97}. For a
fixed equilibrium temperature $T$ as in the present approach, the nonequilibrium evolution according to Eq.\,(\ref{bose})
pushes a certain fraction of bosons into the condensed state for sufficiently large times, which may however not be reached during an actual physical process.


Based on  Eq.\,(\ref{bose}), the equilibration process is driven by elastic collisions that conserve the
total particle number and the particle content of the initial conditions $n_\text{i}(\epsilon)$ is required to agree with
the one of the asymptotic distribution.
However, for large times condensation may set in and consequently
the total particle number at $\epsilon\rightarrow 0$ is expected to have not only a thermal fraction 
$n_\text{eq}(0)< 1/(z¥-1)$ with $z=\exp(-\mu/T)$ and $n_\text{eq}(\epsilon)=1/[z\exp(\epsilon/T)-1]$, but also a condensed fraction $n_\text{c}(t)$ at $\epsilon=0$
which is indirectly accounted for in the nonlinear framework because the total particle number in
the equilibrating system, and in the condensate is conserved.

The particle-number conserving properties of Eq.\,(\ref{bose}) differ from those of the Boltzmann equation with energy-conserving $\delta$-function.
The latter conserves particle number when integrated over $\epsilon \ge 0$ and hence, cannot lead to the formation of a 
condensate, unless there exists one in the initial conditions through an additional $\delta(0)$-function. Since 
Eq.\,(\ref{bose}) emerges from a gradient expansion it allows the system to move into the condensate
without an additional $\delta(0)$-function in the initial conditions, and hence, the thermal solution for $\epsilon \ge 0$
alone does not conserve particle number.
\begin{figure}
	\centering
	\includegraphics[scale=0.6]{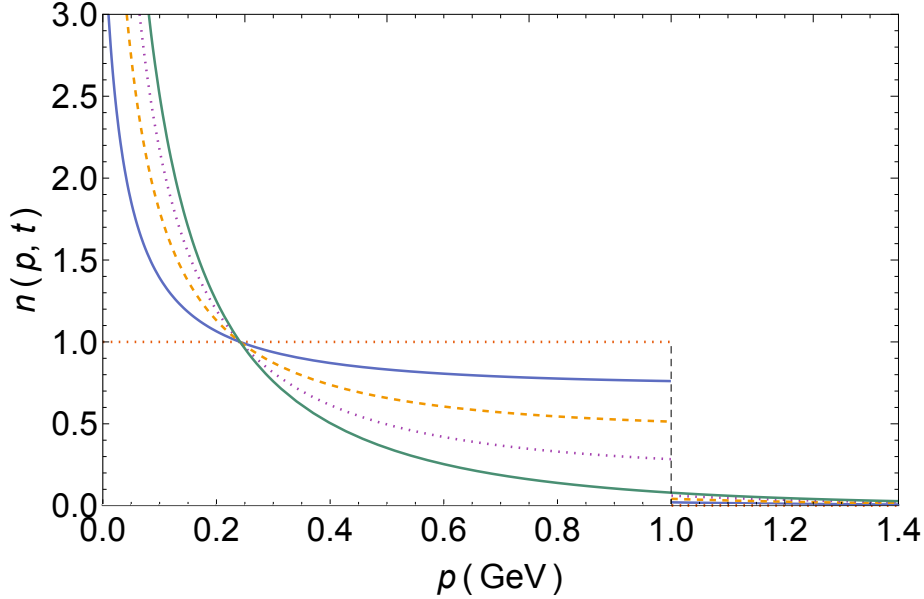}
	\caption{\label{fig1} (color online)  Relaxation of a finite Bose system with initial condition $n_\text{i}(\epsilon)$ from Eq.\,(\ref{ini})  towards the thermal equilibrium distribution $n_\text{eq}(|\bm{p}|)=n_\text{eq}(p)=n_\text{eq}(\epsilon)$ according to the linear Eq.\,(\ref{rela}). The transport coefficients are $D=0.21\times10^{23}$\,GeV$^2$\,s$^{-1},~\,v=-0.53\times10^{23}$\,GeV$\,$s$^{-1}$, the temperature is $T=-D/v\simeq 0.4$\,GeV.
 Times are
	(in units of $10^{-23}$\,s, top to bottom at $p<Q_\text{s}$): 0.1, 0.25, 0.5 and $\infty$. The relaxation time is $\tau_\text{eq}=4D/(9v^2)=0.33\times10^{-23}$\,s $\simeq 1$\,fm/$c$. 
	}
	\end{figure}
	 	\begin{figure}
	\centering
	\includegraphics[scale=0.6]{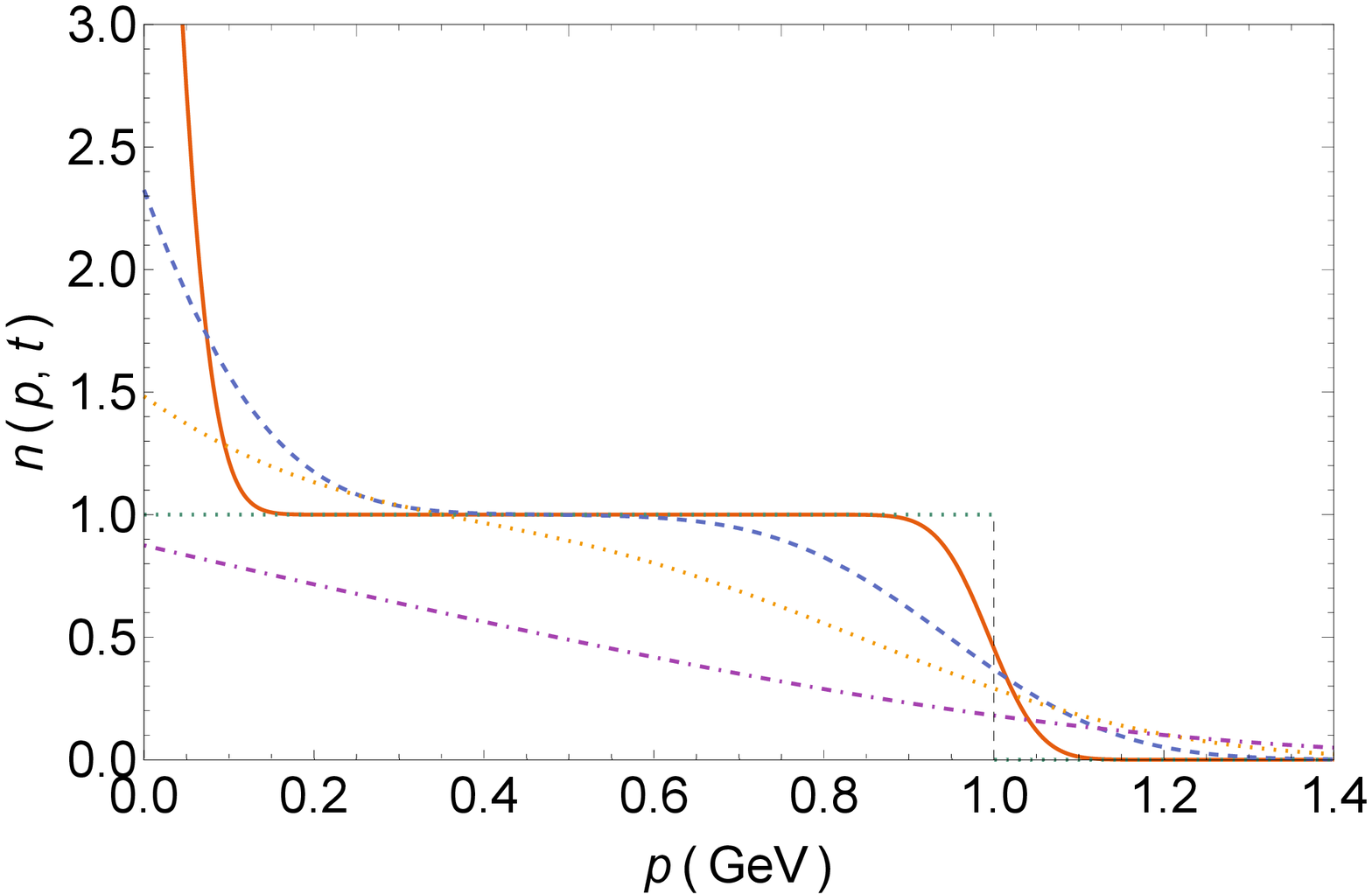}
	\caption{\label{fig2} (color online)  Equilibration of a finite Bose system based on the nonlinear evolution according to Eq.\,(\ref{bose}), but with the $x$-integration in Eq.\,(\ref{nlsolution}) restricted to $x\ge 0$. The transport coefficients $D, v$ with the temperature $T=0.4$\,GeV are as in the linear case in Fig.\,\ref{fig1}. 
	The initial distribution is $n_\text{i}$, box. The solutions $n\,(\epsilon,t)$ of the evolution equation are shown at four values of time $t$ in units of $10^{-23}$\,s: $0.005$ (solid),
$0.05$\,(dashed), $0.15$\,(dotted), $0.5$\,(dash-dotted).
The relaxation time is $\tau_\text{eq}=0.33\times10^{-23}$\,s. In the infrared, the occupation at very short times rises above the thermal limit but is depleted with time and redistributed into the 
	BEC ground state, 
with the thermal occupation $n\,(0,t)<1$ for times $t>\tau_\text{eq}$.
	}
	\end{figure}

	 	\begin{figure}
	\centering
	\includegraphics[scale=0.6]{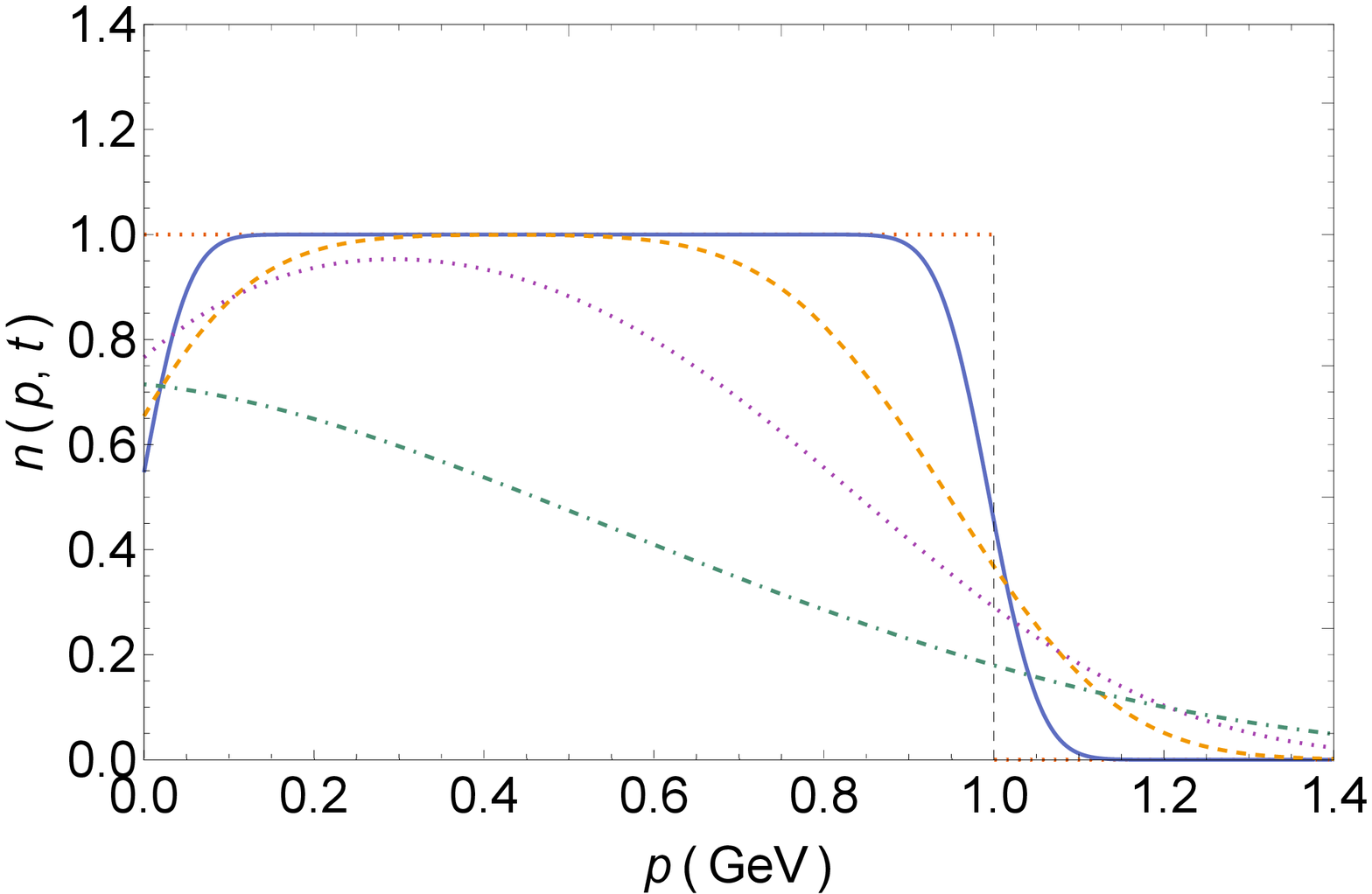}
	\caption{\label{fig3} (color online)  Equilibration of a finite Bose system based on the nonlinear evolution according to Eq.\,(\ref{bose}) with the $x$-integration in Eq.\,(\ref{nlsolution}) extending over the full space $-\infty \le x\le \infty$.  The transport coefficients $D, v$ with the temperature $T=0.4$\,GeV are as in the linear case in Fig.\,\ref{fig1}. The initial distribution is $n_\text{i}$, box. The solutions $n\,(\epsilon,t)$ of the evolution equation are shown at four values of time $t$ in units of $10^{-23}$\,s: $0.005$ (solid),
$0.05$\,(dashed), $0.15$\,(dotted), $0.5$\,(dash-dotted).
The equilibration time is $\tau_\text{eq}=0.33\times10^{-23}$\,s. In the infrared, the occupation is depleted with time and redistributed into the 
	BEC ground state. In the ultraviolet, new a thermal tail develops within $\tau_\text{eq}$.
	}
	\end{figure}

\section{Linear relaxation ansatz}
For a given initial distribution $n_\text{i}(\epsilon)$, an approximate solution of the equilibration problem
can be obtained through the linear relaxation ansatz $\partial\,n_\text{rel}/\partial t=(n_\text{eq} -n_\text{rel})/\tau_\text{eq}$ with the solution
\begin{equation}
n_\text{rel}(\epsilon,t)=n_\text{i}(\epsilon)\,e^{-t/\tau_\text{eq}}+n_\text{eq}(\epsilon)(1-e^{-t/\tau_\text{eq}})\,,
 \label{rela}
\end{equation}
which enforces equilibration towards the thermal distribution $n_\text{eq}(\epsilon)$.
Here, I use the equilibration time $\tau_\text{eq}=4D/(9v^2)$, which will be motivated later 
from the structure of the exact solution of Eq.\,(\ref{bose}). This simplified model will be used to compare with the exact solution
of the nonlinear problem.

For massless bosons, an initial-momentum distribution $n_\text{i}(|\bm{p}|)\equiv\,n_\text{i}(p)=n_\text{i}(\epsilon)$ has been proposed in \cite{mue00} based on \cite{mlv94}.
It accounts, in particular, for the early stages of a relativistic heavy-ion collision \cite{jpb12} assuming that all gluons 
up to a limiting momentum $Q_\text{s}$ are freed on a short time scale
$\tau_0\sim Q_\text{s}^{-1}$ whereas all gluons beyond $Q_\text{s}$ are not freed.
Thus the initial distribution in a volume $V$ is taken to be a constant up to 
$Q_\text{s}$
\begin{equation}
n_\text{i}(\epsilon)=N_\text{i}\,\theta(1-\epsilon/Q_\text{s})\,\theta(\epsilon)\,.
 \label{ini}
\end{equation}
In the corresponding fermionic case \cite{gw82}, all negative-energy states are filled as in the Dirac sea.
Typical gluon saturation momenta for a longitudinal
momentum fraction carried by the gluon $x\simeq 0.01$ turn out to be of the order $Q_\text{s}\simeq 1$\,GeV  \cite{mtw09}, which is chosen for the present model investigation. 

Results for the gluon relaxation from $n_\text{i}(\epsilon)\equiv n_\text{i}(p)$ 
to $n_\text{eq}(p)$ according to Eq.\,(\ref{rela})  are shown in Fig.\,\ref{fig1} for various times.
Thermal equilibrium is reached quickly, within $t\simeq 2\times \tau_\text{eq}$ the distribution function is nearly 
indistinguishable from the equilibrium distribution $n_\text{eq}(p)$, which is approached from below in the infrared region $p\le T\,\log{(2/z)} \simeq 0.24$\,GeV, from above in the UV up to $Q_\text{s}$, and again from below above $Q_\text{s}$. Both $n_\text{i}(\epsilon)$ and $n_\text{eq}(\epsilon)$ are also solutions of Eq.\,(\ref{bose}), but
this is not the case for $n_\text{rel}(\epsilon,t)$. 

The  value $N_\text{i}=1$ is chosen here to allow for a direct comparison with the analytical solution of Eq.\,(\ref{bose}) that will be derived in the next section.
It corresponds to an overoccupied system \cite{jpb12} because the particle number in the initial distribution $N_\text{i}= (4/3)\,\pi\,V Q_s^3\times n_\text{i}$ is larger than
the particle number that can be accommodated at a given temperature $T$ by the thermal equilibrium distribution,
\begin{equation}
N_\text{eq}=4\pi V\int_0^\infty n_\text{eq}(\epsilon)\epsilon^2 d\epsilon
 \label{ninfty}
\end{equation}
for any value of the chemical potential $\mu$. 
If overall particle-number is conserved, the excess particles would be driven into the Bose condensate.

The relaxation ansatz does conserve
the integral over the distribution function, $\int_0^\infty n_\text{i}(\epsilon) d\epsilon = \int_0^\infty n_\text{eq}(\epsilon) d\epsilon= \int_0^\infty n_\text{rel}(\epsilon,t) d\epsilon=T\,\log\,[z/(z-1)]=n_\text{i}\,Q_s~ \forall\,t$.
This is a consequence of the enforced equilibration towards the thermal solution with an adjusted value of $z=1/[1-\exp(-N_\text{i} Q_s/T)]$, or $\mu=-T\log{z}$, for a given temperature $T$.

It cannot be expected that the solution of the nonlinear problem fulfillls a corresponding conservation law.
 Like the relaxation ansatz in the overoccupied case,  Eq.\,(\ref{bose}) will
also not conserve particle number in the nonequilibrium thermal distribution because at large times particles move into the condensate.
Hence it is only the total particle number in the thermal cloud plus the condensate that is conserved.


\section{Exact solution of the nonlinear equation}
 
Whereas nonlinear partial differential equations are rarely solvable in closed form,
in case of Eq.\,(\ref{bose}) an analytical solution can be obtained using a method that 
was proposed in \cite{gw82} for a finite fermion system. Although the approach is similar for bosons,
the different quantum-statistical properties require a new investigation, in particular
regarding the 
transition to a Bose--Einstein condensate (BEC) in a boson
 system.
The transformation
\begin{equation}
n\,(\epsilon,t)=-\frac{D}{vP(\epsilon,t)}\frac{\partial P(\epsilon,t)}{\partial \epsilon}
 \label{nlinear}
\end{equation}
reduces the nonlinear boson Eq.\,(\ref{bose}) to a linear diffusion equation for $P(\epsilon,t)$
\begin{equation}
P_t=-vP_{\epsilon}+DP_{\epsilon\epsilon} 
\label{diffusion}
\end{equation}
where $P_t=\partial P/\partial t, P_{\epsilon}=\partial P/\partial\epsilon.$ 
An equivalent solution of Eq.\,(\ref{bose}) is possible through the linear transformation
\begin{equation}
n\,(\epsilon,t)=\frac{1}{2v}w\,(\epsilon,t)-\frac{1}{2}
 \label{linear}
\end{equation}
which yields Burgers' equation \cite{bur48}  
\begin{equation}
w_t+ww_{\epsilon}=Dw_{\epsilon\epsilon}.
 \label{burgers}
\end{equation}
This equation has the structure of a one-dimensional Navier-Stokes equation without pressure term. It has been 
used to describe fluid flow and, in particular, shock waves in a viscous fluid. It 
can be solved through Hopf's transformation \cite{ho50}
\begin{equation}
w\,(\epsilon,t)=-2D\phi_{\epsilon}/\phi
 \label{hopf}
\end{equation}
which reduces Eq.\,(\ref{burgers}) to the heat equation $\phi_t=D\phi_{\epsilon\epsilon}$. 

The resulting solution of  Eq.\,(\ref{bose}) can then be written as
\begin{equation}
n\,(\epsilon,t)=\frac{1}{2v}\frac{\int_{-\infty}^{+\infty}\frac{\epsilon-x}{t}\,F(\epsilon-x,t)G(x)\,dx}{\int_{-\infty}^{+\infty}F(\epsilon-x,t)G(x)\,dx}-\frac{1}{2}
 \label{nlsolution}
\end{equation}
with a gaussian part arising through the linear diffusion (or heat) equation that extends over the full range of the $x$-integral
from $-\infty$ to $+\infty$
\begin{equation}
F(\epsilon-x,t)=\exp\biggl[-\frac{(\epsilon-x)^2}{4Dt}\biggr]
 \label{feq}
\end{equation}
and an exponential function containing the integral over the initial condition $n_\text{i}(\epsilon)$
\begin{equation}
G(x)=\exp\biggl[-\frac{1}{2D}\bigl(vx+2v\int_0^x n_\text{i}(y)\,dy\bigr)\biggr]\,. 
\label{geq}
\end{equation}\\
The solution Eq.\,(\ref{nlsolution}) can be evaluated for any given nonequilibrium initial distribution,
splitting the integrals conveniently according to eventual discontinuities in the initial conditions. 

The $x$-integration in Eq.\,(\ref{nlsolution}) is over the full energy domain $-\infty \le x \le \infty$.
For bosons, $n_\text{i}(\epsilon<0)=0$, whereas for fermions \cite{gw82}, all negative-energy
states in the Dirac sea are occupied such that due to Pauli's principle the diffusion cannot reach into the
negative-energy region. 

As a consequence of bosons being able to diffuse into the unphysical
domain, Eq.\,(\ref{bose}) conserves particle number only if integrated from $-\infty$ to $+\infty$.
The natural physical interpretation of this behavior is that all particles that move
into the $x \le 0$ region assemble in the condensate, thus conserving the total particle number:
This may be considered as a bosonic equivalent of Pauli's principle.

With $n_\text{i}$ given by Eq.\,(\ref{ini}), I obtain the solution of Eq.\,(\ref{nlsolution})
 either through a transformation of variables,
or using a numerical integration. 
The corresponding results agree, but the analytical approach
is faster and more suitable in view of further conclusions such as the derivation of an explicit expression for the equilibration time.

The exact analytical solution\footnote{Computer algebra programs fail to generate the analytical solution of Eq.\,(\ref{bose}).} of Eq.\,(\ref{bose}) using the initial condition Eq.\,(\ref{ini}) and $N_\text{i}=1$  is
\begin{equation}
n\,(\epsilon,t)=\frac{1}{2v}\times\Bigl[\frac{n_a(\epsilon,t)+n_b(\epsilon,t)+n_c(\epsilon,t)}{n_d(\epsilon,t)+n_e(\epsilon,t)+n_f(\epsilon,t)}\Bigr]-\frac{1}{2}\qquad
\label{exsol}
\end{equation}
with
\begin{eqnarray}
n_a(\epsilon,t)=\exp\Bigl[\frac{1}{2D}(v^2t/2-v\epsilon)\Bigr]\times\Biggl[v\,\sqrt{\pi Dt}\, \Bigl[1+\text{erf}\,(u_0(\epsilon,t))\Bigr]\nonumber
+2D\exp[-(u_0(\epsilon,t))^2]\Biggr],\qquad~~~\\\nonumber
n_b(\epsilon,t)=\exp\Bigl[\frac{1}{2D}(9v^2t/2-3v\epsilon)\Bigr]\times\Biggl[3v\sqrt{\pi Dt}\, \Bigl[\text{erf}\,(u_2(\epsilon,t))-\text{erf}\,(u_1(\epsilon,t))\Bigr]\qquad\qquad\qquad\qquad\nonumber\\
+2D\Bigl[\exp[-(u_2(\epsilon,t))^2]-\exp[-(u_1(\epsilon,t))^2]\Bigr]\Biggr],\qquad\qquad\nonumber\\
n_c(\epsilon,t)=\exp\Bigl[\frac{1}{2D}(v^2t/2-v\epsilon-2v\,Q_s)\Bigr]\times\Biggl[v\,\sqrt{\pi Dt}\, \Bigl[1-\text{erf}\,(u_3(\epsilon,t))\Bigr]\nonumber\qquad\qquad\qquad\qquad\qquad\\
-2D\exp[-(u_3(\epsilon,t))^2]\Biggr],\qquad\nonumber\\
n_d(\epsilon,t)=\sqrt{\pi Dt}\,\exp\Bigl[\frac{1}{2D}(v^2t/2-v\epsilon)\Bigr]\times \Bigl[1+\text{erf}\,(u_0(\epsilon,t))\Bigr]\,,\qquad\qquad\qquad\qquad\qquad\qquad\qquad \nonumber\\
n_e(\epsilon,t)=\sqrt{\pi Dt}\,\exp\Bigl[\frac{1}{2D}(9v^2t/2-3v\epsilon)\Bigr]\times \Bigl[\text{erf}\,(u_2(\epsilon,t))-\text{erf}\,(u_1(\epsilon,t))\Bigr]\,,\qquad\qquad\qquad\qquad~~  \nonumber\\
n_f(\epsilon,t)=\sqrt{\pi Dt}\,\exp\Bigl[\frac{1}{2D}(v^2t/2-v\epsilon-2v\,Q_s)\Bigr]\times\Bigl[1-\text{erf}\,(u_3(\epsilon,t))\Bigr]\,.\qquad\qquad\qquad\qquad\qquad~ \nonumber
\end{eqnarray}
The auxiliary functions $u_i\,(\epsilon,t)$ are
\begin{eqnarray}
u_0(\epsilon,t)=\frac{1}{2\sqrt{Dt}}(-\epsilon+vt)\,,\qquad\qquad\qquad\qquad\qquad\qquad\qquad\nonumber\\
u_1(\epsilon,t)=\frac{1}{2\sqrt{Dt}}(-\epsilon+3vt)\,,\qquad\qquad\qquad\qquad\qquad\qquad\qquad\nonumber\\
u_2(\epsilon,t)=\frac{1}{2\sqrt{Dt}}(Q_s-\epsilon+3vt)\,,\qquad\qquad\qquad\qquad\qquad\qquad\nonumber\\ 
u_3(\epsilon,t)=\frac{1}{2\sqrt{Dt}}(Q_s-\epsilon+vt)\,.\qquad\qquad\qquad\qquad\qquad\qquad~ \nonumber
\label{solution}
\end{eqnarray}

One may argue that the $x$-integration in the solution Eq.\,(\ref{nlsolution})} of the nonlinear equation should be restricted to $x \ge 0$ since physically meaningful solutions 
cover only the positive semi-axis. In the analytical solution Eq.\,(\ref{exsol})} this corresponds to omitting $n_a(\epsilon,t)$ and $n_d(\epsilon,t)$, thus yielding the result
shown in Fig.\,\ref{fig2}.
Here the nonlinear evolution is seen to indeed remove the discontinuities seen in the relaxation solutions (Fig.\,\ref{fig1}) at $\epsilon=Q_\text{s}$. 

At short times, however, the solutions display 
a simultaneous rise of occupation in the IR that is initially far above the thermal value $n_\text{eq}(0)=1/(z-1)$\,(with $n_\text{eq}(0)\simeq11.5$ for the example shown
 in Fig.\,\ref{fig2} where $z=1.087$), but then drops rapidly, reaching again $n(\epsilon\simeq 0,t)=1$ for
times $t\simeq \tau_\text{eq}$. 
This change in the time evolution
corresponds to a change of sign in the argument of the error functions 
 for $t\sim\tau_\text{c}(\epsilon)=1/[v(1+2\,N_\text{i})]\times(\epsilon-Q_\text{s})$.

However, if the full space is considered without restricting the $x$-integration to the positive semi-axis as is appropriate for the general solution
of the Burgers' part \cite{bur48,ho50} of the nonlinear equation Eq.\,(\ref{bose})}, the behavior of the solutions changes as displayed in 
Fig.\,\ref{fig3}. Whereas in the UV, the nonlinear evolution still removes the discontinuities seen in the relaxation solutions at $\epsilon=Q_\text{s}$,
in the IR the solutions show a simultaneous fall of occupation below the initial value.
It occurs because the bosons in the infrared region diffuse into the $\epsilon \le 0$ domain, they are not any more part of the
thermal cloud. 

The physical interpretation of this feature
 characterizing the nonlinear evolution for bosons is that condensation has occurred even though no seed condensate had been specified
in the initial conditions (as would have been required in the Boltzmann equation with a $\delta$-function for the conservation of energy).
The full solution thus exhibits
 the expected physical behavior in both the IR and UV regions, and it fulfillls 
$n\,(\epsilon,t)\rightarrow n_\text{i}(\epsilon)$ for $t\rightarrow 0$.

\begin{figure}
	\centering
	\includegraphics[scale=0.6]{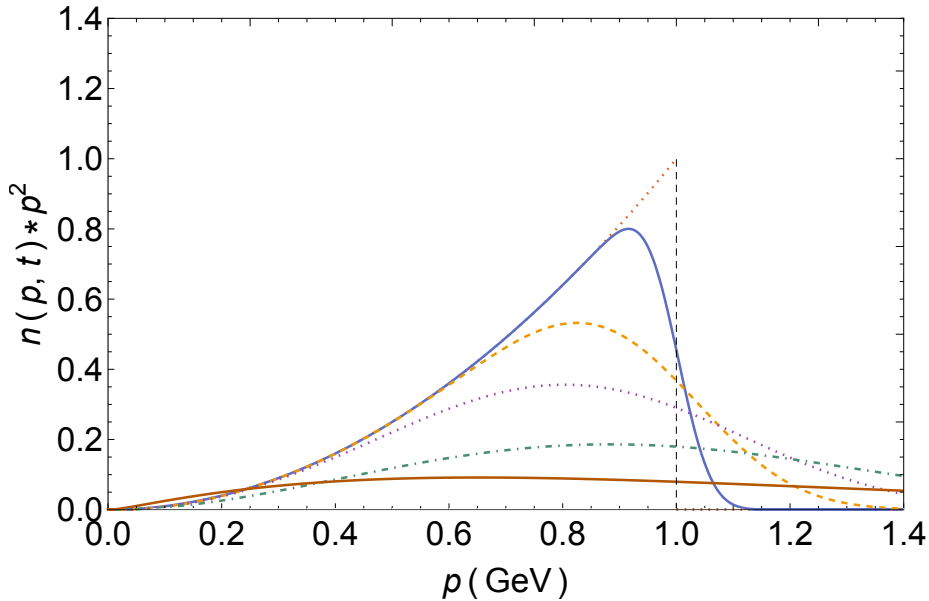}
	\caption{\label{fig4} (color online)  Integrands $n(\epsilon=p,t)\,p^2$ for the nonlinear evolution according to Eq.\,(\ref{bose}) as in Fig.\,\ref{fig3}.
The initial distribution for $N_\text{i}=1$ (overoccupation) is $n_\text{i}\,(p)\,p^2$\,(upper curve, dotted). The integrands $n(p,t)\,p^2$ 
are shown for four values of time $t$ in units of $10^{-23}$\,s: 
$0.005$\,(solid), $0.05$\,(dashed), $0.15$\,(dotted), $0.5$\,(dash-dotted), from top to bottom. The corresponding  thermal equilibrium solution $n_\text{eq}(p)\,p^2$ is displayed as solid curve, bottom. 
Particle number is not conserved during the time evolution since excess particles move into the condensate.
	}
	\end{figure}

Hence, in the course of the time evolution the particle content in the initial distribution function is depleted, and for conserved total particle number it reappears
as occupation of the condensate $N_\text{c}(t)=(2\pi)^3\,n_\text{c}(t)\times\delta(\bm{p})$, as expected for a Bose system. The
integral $N_\text{th}(t)=\int_0^\infty n\,(\epsilon,t)\,\epsilon^2 d\epsilon$ is not conserved, as was also the case for the relaxation ansatz in the
overoccupied situation. This is confirmed by a numerical integration of $n\,(\epsilon,t)\,\epsilon^2$ as displayed
in Fig.\,\ref{fig4}. 

The total particle number $N_\text{th}(t)+N_\text{c}(t)$, however, is conserved. This can be verified by integrating the analytical
time-dependent distribution functions over the full energy space 
\begin{equation}
\label{nbose}
N_\text{tot}(t)=\int_{-\infty}^\infty n(\epsilon,t)\,g\,(\epsilon)\,d\epsilon=\text{const}\,.
\end{equation}
 I have checked this relation through a numerical integration for constant density of states. (For $g\,(\epsilon)\ne$ const, $n_\text{i}(\epsilon)$ must first be renormalized such that $\int_0^\infty n_\text{i}(\epsilon)\,g\,(\epsilon)\,d\epsilon=
 \int_0^\infty n_\text{eq}(\epsilon)\,g\,(\epsilon)\,d\epsilon$
 in order to conserve particle number, with $n(\epsilon,t)$ recalculated for $N_\text{i}<1$). 
 
 Changing the integration range from the positive semi-axis (as in the Boltzmann equation with energy-conserving $\delta$-function)
 to the full energy range may seem arbitrary, but it is a necessary consequence of the diffusion equation that occurs in the solution of the nonlinear Eq.\,(\ref{bose}) since diffusion is not restricted to the positive semi-axis and hence, particle-number conservation is only fulfillled if the integration is over the full space.

An explicit expression for the equilibration time in the case of initial occupation $N_\text{i}=1$ follows from an asymptotic expansion of the error functions with arguments $z_\text{b}$ occurring in the solution at the boundary $x_\text{b}=Q_\text{s}$, 
\begin{eqnarray}
\text{erf}\,(z_\text{b})=\text{erf}\biggl[\frac{1}{2\sqrt{Dt}}(x_\text{b}-\epsilon+3vt)\biggr]\simeq\qquad\qquad\qquad\nonumber \\\\
\simeq1-\frac{2\sqrt{Dt}}{\sqrt{\pi}\,(x_\text{b}-\epsilon+3vt)}\nonumber
\exp\biggl[-\frac{1}{4Dt}(x_\text{b}-\epsilon+3vt)^2\biggr]+ \exp(-z^2)\mathcal{O}\Bigl(\frac{1}{z_\text{b}^3}\Bigr).~
\label{erf}
\end{eqnarray}\\

For initial occupations different from $N_\text{i}=1$ the argument is modified according to
\begin{equation}
z_\text{b}=\frac{1}{2\sqrt{Dt}}\,(x_\text{b}-\epsilon+vt+2vN_\text{i}\,t)\,.
 \label{arg}
\end{equation}

Therefore, deviations from the asymptotic solution are vanishing with time $\propto \exp\,[-(1+2\,N_\text{i})^2v^2t/(4D)]$
such that
the equilibration time in a system of bosons with initial condition Eq.\,(\ref{ini}) becomes
\begin{equation}
\tau_\text{eq}^\text{Bose}=4D/[(1+2\,N_\text{i})^2\,v^2]\,,
 \label{taubose}
\end{equation}
as was already used before
 in the linear model calculation based on the relaxation ansatz. 
The Bose equilibration time for $N_\text{i}=1$ and transport coefficients $D, v$ with temperature $T=-D/v$ is a factor nine shorter than
the corresponding equilibration time in a fermion system, which was found to be $\tau_\text{eq}^\text{Fermi}=4D/v^2$ in \cite{gw82}.

This difference is solely due to the quantum-statistical properties of a bosonic as compared to a fermionic system:
In a fermion system, changes of the occupation of single-particle states are suppressed due to the exclusion principle
and hence, in case of fixed transport parameters $D, v$ the equilibration process takes more time for fermions than for bosons. 

The equilibration of a Fermi system had been considered in \cite{gw82}. Rewriting Eq.\,(\ref{bose}) for fermions -- substituting $n\,(1+n)$ by $n\,(1-n)$ --,
the solution of the corresponding fermionic equilibration problem\footnote{In Eq.\,(14) of \cite{gw82}, $n_0(x)$
is replaced by $m(x,\epsilon;t)= \frac{1}{2} -\frac{\epsilon - x}{2 v t}$\,.} becomes in analogy to Eq.\,(\ref{nlsolution}) \begin{equation}
n_\text{F}\,(\epsilon,t)=-\frac{1}{2v}\frac{\int_{-\infty}^{+\infty}\frac{\epsilon-x}{t}\,F(\epsilon-x,t)G_\text{F}(x)\,dx}{\int_{-\infty}^{+\infty}F(\epsilon-x,t)G_\text{F}(x)\,dx}+\frac{1}{2}
 \label{fnlsolution}
\end{equation}
with the exponential function $G_\text{F}$ that contains the integral over the fermionic initial conditions
\begin{equation}
G_\text{F}(x)=\exp\biggl[-\frac{1}{2D}\bigl(vx-2v\int_0^x n_\text{i}(y)\,dy\bigr)\biggr]\,. 
\label{fgeq}
\end{equation}\\
The function $G_\text{F}$ differs from the bosonic case (Eq.\,(\ref{geq})) due to the minus sign in front of the last term. The integrals $\int_{-\infty}^{+\infty}$
for fermions can be evaluated analytically or numerically, with initial conditions given by Eq.\,(\ref{ini}) to compare with the bosonic case -- see \cite{gw82} for nonrelativistic energies and other initial conditions. 
For fermions, the initial negative-energy states in the Dirac sea
must be taken as occupied, $n_\text{i}(\epsilon<0)=1$ -- otherwise, at large times $t$ the solutions near $\epsilon\simeq 0$ tend to drop below $n(0,t)=1$ because occupation
drifts into the $\epsilon < 0$ domain.

\begin{figure}
	\centering
	\includegraphics[scale=0.6]{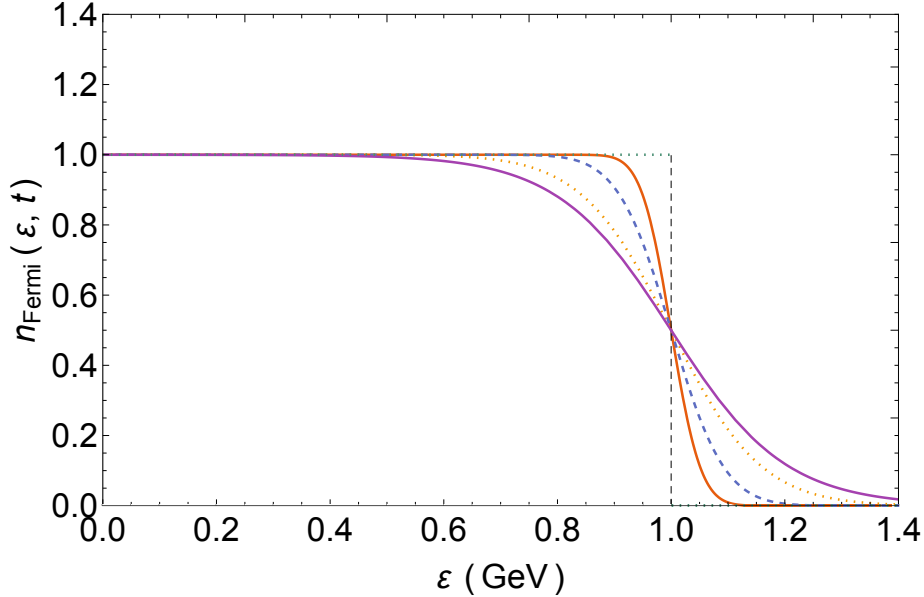}
	\caption{\label{fig5} (color online)  Equilibration of a finite Fermi system based on the nonlinear evolution according to Eq.\,(\ref{fnlsolution}), see also \cite{gw82}. 
	The transport coefficients are $D=0.21\times10^{23}$\,GeV$^2$\,s$^{-1},~\,v=-2.1\times10^{23}$\,GeV$\,$s$^{-1}$, the temperature is $T=-D/v\simeq 0.1$\,GeV.
The equilibration time is $\tau_\text{eq}^\text{Fermi}=4D/(v^2)=0.19\times10^{-23}$\,s $\simeq 0.6$\,fm/$c$. 
The initial distribution is $n_\text{i}\,(\epsilon)$\,(box), the solutions $n_\text{F}(\epsilon,t)$ of the evolution equation are shown at three values of time $t$ in units of $10^{-23}$\,s: $0.005$ (solid),
$0.02$\,(dashed), and $0.1$\,(dotted). The corresponding  thermal equilibrium solution $n_\text{eq}^\text{F}(\epsilon)$ is displayed as solid curve. It is approached asymptotically by the solutions of the nonlinear equation for fermions.
	}
	\end{figure}

The resulting solutions for the fermionic equilibration problem are shown in Fig.\,\ref{fig5}. The system is seen to  approach the thermal equilibrium distribution for $t\rightarrow \infty$ for all energies 
$\epsilon \ge 0$ 
\begin{equation}
n_\text{eq}^\text{F}(\epsilon)=\frac{1}{e^{(\epsilon-Q_s)/T}+1}\,.
 \label{Fermi}
\end{equation}
Here I have rescaled the transport coefficients  as compared to the bosonic case in Fig.\,\ref{fig3}\, 
such that the temperature is $T=0.1$\,GeV, and the equilibration time 
$\tau_\text{eq}^\text{Fermi}=4D/(v^2)=0.19\times10^{-23}$\,s $\simeq 0.6$\,fm/$c$\,. With the smaller temperature as compared to the
bosonic case, the fermionic equilibrium distribution has a rather narrow diffuseness and the Dirac sea at negative energies remains fully occupied.

\begin{figure}
	\centering
	\includegraphics[scale=0.6]{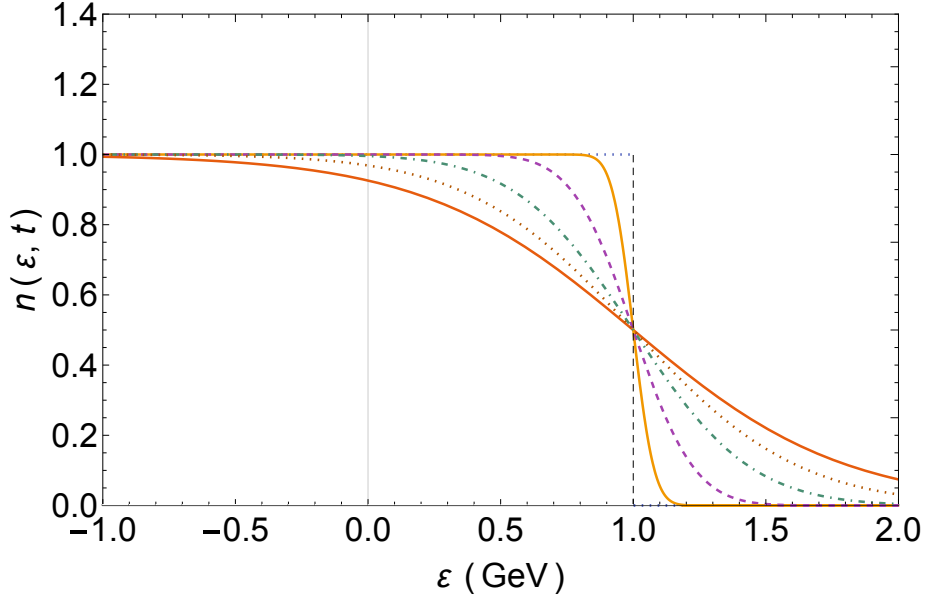}
	\caption{\label{fig6} (color online)  Equilibration of a finite Fermi system based on the nonlinear evolution according to Eq.\,(\ref{fnlsolution}) as in Fig.\,\ref{fig5},
	but with the same transport coefficients as in Fig.\,\ref{fig3} for bosons:
	$D=0.21\times10^{23}$\,GeV$^2$\,s$^{-1},~\,v=-0.53\times10^{23}$\,GeV$\,$s$^{-1}$, the temperature is $T=-D/v\simeq 0.4$\,GeV.
The equilibration time is 9$\times$ larger than for bosons, $\tau_\text{eq}^\text{Fermi}=4D/(v^2)\simeq 3\times10^{-23}$\,s.
The initial distribution is $n_\text{i}\,(\epsilon)$\,(box), the solutions $n_\text{F}(\epsilon,t)$ of the evolution equation are shown at four values of time $t$ in units of $10^{-23}$\,s: $0.01$ (solid),
$0.1$\,(dashed), $0.5$\,(dot-dashed) and $1.5$\,(dotted). The corresponding  thermal equilibrium solution $n_\text{eq}^\text{F}(\epsilon)$ is displayed as solid curve. It is approached asymptotically by the solutions of the nonlinear equation for fermions. For a temperature of about 400 MeV, antiparticles are created from the Dirac sea (negative-energy region) for sufficiently large times. 
	}
	\end{figure}

Using the same transport coefficients $D, v$ as in Fig.\,\ref{fig3} for bosons --  that correspond to a temperature of about 400 MeV -- 
produces very broad fermionic distributions as shown in Fig.\,\ref{fig6}.  For sufficiently large times, these extend into the negative-energy domain,
thus accounting for antiparticle creation from the Dirac sea. Whereas in the Boltzmann equation with energy-conserving $\delta$-function antiparticles have positive energies, due to the gradient expansion and the appearance of the diffusion equation in the solution of the nonlinear equation \cite{gw82} fermionic  antiparticles correspond to holes in the Dirac sea at negative energies. Hence, in order to fulfill particle-number conservation, their distribution must be deduced from the particles at positive energies.

Regardless of the choice of transport coefficients, Pauli's principle confines the fermionic distribution function to $n_\text{F}(\epsilon,t)\leq 1$. Particle number is conserved during the
time evolution provided antiparticle creation is also taken into account (second integral in Eq.\,(\ref{nth}); $g\,(\epsilon)\propto \epsilon^2$ is the density of states for a relativistic system)
\begin{equation}
\label{nth}
N_\text{th}^\text{F}(t)=\int_0^\infty n_\text{F}(\epsilon,t)\,g\,(\epsilon)\,d\epsilon-\int_{-\infty}^0 [1-n_\text{F}(\epsilon,t)]\,g\,(\epsilon)\,d\epsilon=\text{const}\,,
\end{equation}
 and obviously no condensate is formed. Whereas it is easy to test this relation for the thermal equilibrium distribution
 $n_\text{F}(\epsilon,t)\rightarrow n^\text{F}_\text{eq}(\epsilon)$, I have also verified it with high precision in numerical integrations of the time-dependent fermion
 distribution functions $n_\text{F}({\epsilon,t})$ that solve Eq.\,(\ref{fnlsolution}) analytically.
  (For $g\,(\epsilon)\ne$ const, $n_\text{i}(\epsilon)$ must first be renormalized as in the bosonic case such that $\int_0^\infty n_\text{i}(\epsilon)\,g\,(\epsilon)\,d\epsilon=
 \int_0^\infty n_\text{eq}(\epsilon)\,g\,(\epsilon)\,d\epsilon$
 in order to conserve particle number, with $n_\text{F}(\epsilon,t)$ recalculated for $N_\text{i}<1$.) 
	
Since the initial state in relativistic heavy-ion collisions is determined mostly by gluons, the short 
equilibration times encountered there are to a large extent due to the statistical factors for bosons.
It is interesting that $\tau_\text{eq}^\text{Bose}\sim 1/(1+2\,N_\text{i})^2$ decreases with rising occupation $N_\text{i}$ of
the initial state. In particular it has been shown in \cite{jpb12,jpb13,jpb14} that in case of particle-number conservation 
the initial system may contain more gluons than are compatible
with the equilibrium distribution. It is then overoccupied as in the $N_\text{i}=1$ case discussed here, which appears to be a 
necessary condition to move into the condensed state for large times $t>\tau^\text{Bose}_\text{eq}$.


Based on Eq.\,(\ref{bose}), the analytical solution differs from $n_\text{rel}$ in particular in the infrared where it drops towards zero since the particles tend to condense, and then disappear from the distribution. In the UV, the solution of Eq.\,(\ref{bose}) removes the discontinuities that persist in $n_\text{rel}(\epsilon,t)$. The UV equilibration is faster than in the relaxation approach due to the action of the nonlinearity, as is evident from the different time sequences in Fig.\,\ref{fig1} and Fig.\,\ref{fig3}.
	 	\begin{figure}
	\centering
	\includegraphics[scale=0.6]{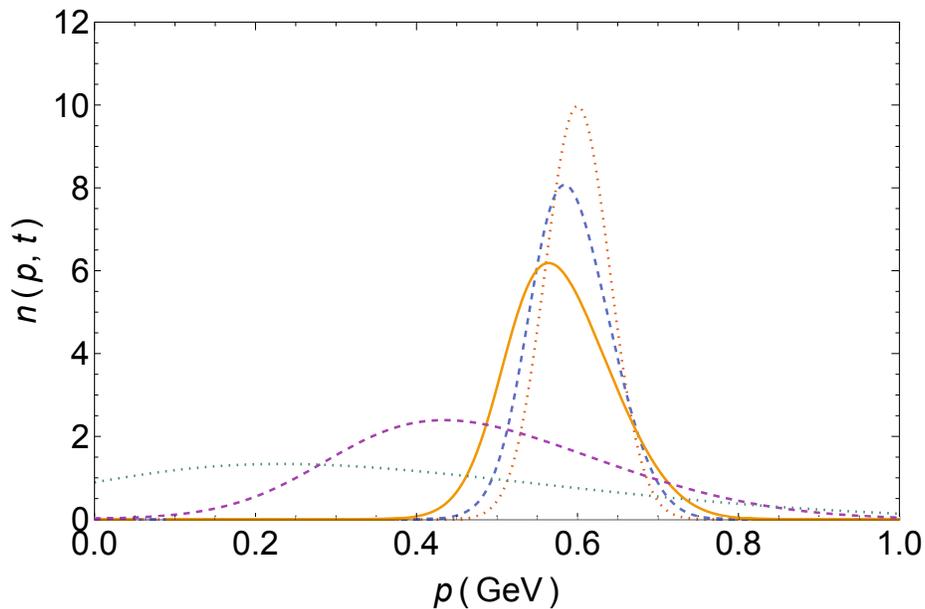}
	\caption{\label{fig7} (color online)  Equilibration of a finite Bose system based on the nonlinear evolution according to Eq.\,(\ref{bose}) for gaussian initial conditions (peaked curve, dotted)
with standard deviation $\sigma=0.04$ GeV.
	 The transport coefficients $D, v$, the temperature $T=0.4$\,GeV and the equilibration time $\tau_\text{eq}=0.33\times10^{-23}$\,s are as in Fig.\,\ref{fig3}.
The solutions $n\,(\epsilon,t)$ of the evolution equation are shown at four values of time $t$ in units of $10^{-23}$\,s: $0.002$ (dashed),
$0.006$\,(solid), $0.06$\,(dashed), $0.2$\,(dotted). 
In the infrared, the occupation 
is redistributed into the BEC ground state. In the UV, a thermal tail develops.
	}
	\end{figure}

Depending on the boson equilibration problem under investigation, it is of interest to solve Eq.\,(\ref{bose}) for 
different initial conditions. As an example, I have tested the gaussian case 
$n_\text{i}(\epsilon)=N_\text{i}\,(\sqrt{2\pi}\sigma)^{-1}\exp\,\bigl[(\epsilon-\langle\epsilon_\text{i}\rangle)/(2\sigma^2)\bigr]$
with $\epsilon_\text{i}=0.6$\,GeV, $\sigma=0.04$\,GeV and $N_\text{i}=1$ using a numerical solution of Eq.\,(\ref{nlsolution}). 

The result is shown in 
Fig.\,\ref{fig7} where the initial distribution $n_\text{i}\,(\epsilon)$ and four solutions $n\,(\epsilon,t)$ of the nonlinear evolution equation are displayed.
With increasing time, the distribution function broadens and its mean value shifts towards the infrared
where condensation eventually sets in.
At times $t\,>\,0.2\times 10^{-23}$\,s,
the distribution function approaches zero rather uniformly across the whole spectrum and eventually at $T<<T_c$ most of the particles would sit in the condensate -- provided the total particle number is conserved.

Whereas for high-energy fermions, the drift towards $p\le 0$
and the diffusion in the infrared tail can
lead to antiparticle formation, in case of bosons with particle-number conservation
a condensate may be formedÑ provided the drift is sufficiently large
such that the temperature $T$
is below the critical value, $T=-(D/v)\times \theta[1+D/(v\,T_c)]$. 

For smaller drift coefficients $v$ corresponding to higher temperatures $T=-D/v$, the shift of the mean value of the distribution towards zero is reduced (or absent for $v\rightarrow0$), and condensate formation would only be possible in the infrared tail of the distribution function due to diffusion. The probability for BEC formation can be calculated for any given initial distribution and fixed values of the transport coefficients that determine the temperature. 

There is, however, presently no microscopic calculation of the transport coefficients available. For a gluonic system, these may eventually be related to the
ratio 
of shear viscosity to entropy density \cite{kov05}. However, since the model uniquely connects $v$ and $D$ 
with $T$ and $\tau_\text{eq}$, fixed values of the latter for $N_\text{i}=1$ have been chosen to compute $D,v$.

\section{Conclusion}

To summarize, I have outlined a schematic model for equilibration in finite Bose systems. The master equation for the occupation numbers 
has been transformed into a nonlinear partial differential equation that keeps track of the statistical factors in an essential way.
In particular, it allows the system
to evolve into the condensate.
A closed-form solution has been obtained.

In the simplified case
of constant transport coefficients and schematic initial conditions that correspond to the ones for a finite gluon system,
\linebreak 
I have derived the exact analytical solution as well as the 
expression for the equilibration time. The distribution function accounts for the depletion of particles
in the course of time because of condensation, and for the simultaneous buildup of a thermal tail in the
UV within the equilibration time.


The decreasing IR populations at late times are counterbalanced by the onset of gluon condensation. 
The transition to the condensate is not
accounted for explicitly in the nonequilibrium model because a Boltzmann-type approach does not
include the required phase coherence.
It is, however, considered indirectly through overall particle-number conservation in the equilibrating system including the condensate.
Hence it appears that the time evolution described by Eq.\,(\ref{bose}) is a realistic approximation of the dynamics of a Bose gas.

Dissipative and diffusive effects combine with the nonlinearity to yield the time evolution towards the asymptotic distribution,
which differs from the usual thermal limit since the system could eventually reach the condensed state at $\epsilon=0$. 
Whether this is actually the case in a heavy-ion collision depends, in particular, on the relevance of inelastic 
number-changing processes, which will most likely prevent the onset of condensation.

Apart from the condensation problem, it is conceivable to apply the time dependence provided by the present analytical model to the gluon system
that is initially created in relativistic collisions at early times $t<\tau_\text{eq}$, thus serving as a starting point for the subsequent evolution of the expanding
quark--gluon plasma in the fireball. One would then have to couple the solution of the gluonic equilibration problem to a dynamic (for example, viscous hydrodynamic)
description of the system which accounts for the spatial collective expansion of the fireball that is believed to occur in relativistic heavy-ion collisions.

Regarding cold quantum gases, a suitably adapted 
model may turn out to be relevant to account for the equilibration in the thermal cloud and the formation of a Bose-Einstein condensate \cite{gw18}.
Indeed applications of  the nonlinear model to non-relativistic cases with finite mass $m$ and dispersion relation $\epsilon=|\bm{p}|^2/(2m)$ 
could be most interesting.
\\\\
\bf{Acknowledgements}
\rm

Discussions with David Blaschke at  the University of Wroc{\l}aw,
Kenji Fukushima
during his stay at the Institute for Theoretical Physics in Heidelberg, and Johannes H\"olck at ITP 
are gratefully acknowledged.
\bibliographystyle{elsarticle-num}
\bibliography{gw17}
\end{document}